\documentclass[10pt,aps,pre,twocolumn,floatfix,notitlepage,superscriptaddress,preprintnumbers,longbibliography]{revtex4-1}

\usepackage{amsmath,amssymb,amsfonts}
\usepackage{graphicx,color}
\usepackage{subfigure}
\usepackage{bm}
\usepackage{verbatim}
\usepackage{float}
\usepackage{dcolumn}
\usepackage{natbib}
\usepackage{hyperref}
\usepackage{algorithm}
\usepackage{algpseudocode}

\makeatletter
\def\BState{\State\hskip-\ALG@thistlm}
\makeatother

\newcommand{\ie}{\emph{i.e.}}
\newcommand{\eg}{\emph{e.g.}}

\begin{document}
\title{Generalized Markov stability of network communities}

\author{Aurelio Patelli}
\affiliation{Istituto dei Sistemi Complessi (CNR) UoS Dipartimento di Fisica, ``Sapienza'' Universit\`a di Roma, 00185 Rome (Italy)}
\affiliation{Service de Physique de l'Etat Condens\'e, UMR 3680 CEA-CNRS, Universit\'e Paris-Saclay, CEA-Saclay, 91191 Gif-sur-Yvette (France)}
\author{Andrea Gabrielli}
\affiliation{Dipartimento di Ingegneria, Universit\`a Roma 3, 00146 Rome (Italy)}
\affiliation{Istituto dei Sistemi Complessi (CNR) UoS Dipartimento di Fisica, ``Sapienza'' Universit\`a di Roma, 00185 Rome (Italy)}
\author{Giulio Cimini}
\affiliation{Dipartimento di Fisica, Universit\`a di Roma Tor Vergata, 00133 Rome (Italy)}
\affiliation{Istituto dei Sistemi Complessi (CNR) UoS Dipartimento di Fisica, ``Sapienza'' Universit\`a di Roma, 00185 Rome (Italy)}

\begin{abstract}
We address the problem of community detection in networks by introducing a general definition of Markov stability, based on the difference between the probability fluxes of a Markov chain on the network at different time scales. 
The specific implementation of the quality function and the resulting optimal community structure thus become dependent both on the type of Markov process and on the specific Markov times considered. 
For instance, if we use a natural Markov chain dynamics and discount its stationary distribution -- that is, we take as {\em reference process} the dynamics at infinite time -- we obtain the standard formulation of the Markov stability. 
Notably, the possibility to use finite-time transition probabilities to define the reference process naturally allows detecting communities at different resolutions, without the need to consider a continuous-time Markov chain in the small time limit. 
The main advantage of our general formulation of Markov stability based on dynamical flows is that we work with lumped Markov chains on network partitions, having the same stationary distribution of the original process. 
In this way the form of the quality function becomes invariant under partitioning, leading to a self-consistent definition of community structures at different aggregation scales. 
\end{abstract}
\maketitle

\section{Introduction}
\label{sec:introduction}

Networks are systems made up of entities (nodes) embedded in a complex pattern of interconnections (links), which occur in a large variety of contexts -- ranging from socio-economic systems and infrastructures to biological processes and ecosystems \cite{RevModPhys.74.47,BOCCALETTI2006175,RevModPhys.80.1275,Cimini2019review}. Networks observed in nature have a recurrent set of characteristics, such as fat-tail behavior of the degree distribution, small-world topology, and community structure 
-- the latter referring to the internal organization of nodes into densely connected groups. 
Identifying the communities of a network means uncovering its mesoscopic structure, and is still an outstanding challenge for network science \cite{Fortunato2010,Fortunato2016,Schaub2017}.


The first method proposed in the literature to partition a network in communities is based on the maximization of a quality function, the {\em modularity}, which compares the actual number of links in the network falling inside each community to the expectation of such number under a null network model \cite{Girvan2002}. 
The modularity function has been then generalized to various setups, like directed, weighted or bipartite networks (see \eg~\cite{Newman2003,Lancichinetti2011}), and still nowadays represents the benchmark method for community detection~\cite{Fortunato2016}. 
However, by relying on a global null model, the modularity suffers from a resolution limit, that is, it cannot find communities smaller than a minimum size -- which depends on the scale of the whole system~\cite{Fortunato2007}. 
Multi-resolution versions of the modularity address this issue using a tunable resolution parameter \cite{Arenas2008,Reichardt2006}, whereas, the modularity-density functional employs a penalty function for splitting partitions~\cite{Chen2015}.
Another popular approach to community detection consists in fitting the network to a stochastic blockmodel, namely a random graph with built-in communities \cite{Karrer2011}, yet this approach was recently shown to be equivalent to modularity maximization \cite{Newman2016a}.
Other well known community detection methods use clique percolation~\cite{Palla2005}, spectral graph properties~\cite{Newman2006}, spin glass models~\cite{Reichardt2006,Traag2011,Traag2013} or combinatorial arguments -- notably this latter method, {\em Surprise}  \cite{Aldecoa2011,Aldecoa2013}, is nearly unaffected by the resolution limit, but has the opposite drawback of overestimating the number of communities \cite{Traag2015}.

Another popular branch of community detection methods is based on {\em random walks}~\cite{MASUDA20171}. 
The idea is that communities correspond to network regions where the walker's dynamics spends a relatively long time, because of the high density of links within communities and the sparse connections across communities. 
This phenomenon leads to the definition of a quality function known as {\em Markov stability} \cite{Delvenne2010}. 
Notably, Markov stability allows interpolating between modularity and spectral clustering by simply varying the time scale of the dynamics~\cite{Lambiotte2014}.
Indeed, such a time scale effectively acts as a resolution parameter, with short scales leading to many small communities and long scales to a few large communities \cite{Delvenne2010,Kheirkhahzadeh2016}. 
Using continuous-time random walks in the small time limit can even overcome the resolution limit of the modularity \cite{Delvenne2010}.  
Among related methods, the {\em Walktrap} algorithm has been one of the first to use random walks for inferring similarities between nodes whence the network community structure~\cite{Pons2006}. 
The popular {\em Infomap} algorithm instead puts the community detection problem in information-theoretical terms \cite{Rosvall2008,Rosvall2009}: 
the functional to be optimized with respect to the network partition is the description length for the moves of a random walker on the network. 
Hence the codebook and the codewords are based on the transition probabilities and stationary distribution of the random walk. 
Related to this, methods based on Boltzmann minimum description length have recently been proposed \cite{Perotti2018}. 
Random walks have also been used to partition the links (rather than the nodes) of the network, and thus to uncover community structures using the concept of the line graph \cite{Evans2009,gabrielli2019}.

The plethora of community detection methods give similar but not identical results, and indeed no algorithm seems to be optimal for all possible community detection tasks \cite{Aldecoa2013b,Peel2017}. 
This happens because community detection is an ill-defined problem: there is no universal definition of communities, and thus no clear guidelines on how to build and assess a community detection method \cite{Fortunato2016,Schaub2017}. 
For instance, the approaches based on the network topology (modularity and blockmodel) or on link combinatorics (surprise) use a null network model to assess the statistical significance of a network partition, and the freedom in choosing the null model introduces a degeneracy in the definition of the community structure.
Physics-inspired methods suffer from the same pathology, since changing the definition of the interaction between nodes and the strength of the noise give different phases, 
whereas, methods based on random walks find different partitions depending on the particular dynamics implemented on the network \cite{Zlatic2010}.

Given that the quest for the best method to detect the ``true'' communities of any network is possibly vain, here we follow up on the complementary viewpoint of random walks methods that any given dynamical process on the network induces a different community structure \cite{Lambiotte2014}. 
We thus consider a general Markov diffusion process on the network, and derive a general quality function for the optimization problem using the transition probability fluxes of the dynamics at different time scales. 
In this way we generalize previous definitions of the Markov stability, which compare the Markov dynamics at finite times to a reference process given by its stationary distribution (\ie, the dynamics at infinite time) \cite{Delvenne2010,Lambiotte2014}. 
Indeed by varying the time scales of the Markov dynamics and of the reference process we can detect communities at both higher and lower resolutions. 
Remarkably, our approach is grounded on the definition of lumped Markov chains on network partitions \cite{Piccardi2011}, whose stationary distributions follow the same aggregating rules of the dynamics. Thanks to this property the form of the quality function becomes invariant under network partitioning, leading to a self-consistent definition of communities at different aggregation scales. 
This leads not only to an elegant theoretical formulation of the problem but also to a convenient recursive algorithm for the optimization of the quality function.

\section{Markov chain on networks}
\label{sec:MC}

We start by recalling basic definitions and properties of Markov chains on networks. We then introduce lumped Markov chains on network partitions, and illustrate these concepts in the simple case of the natural Markov chain (\ie, the random walk).

A network is a set $\mathcal{N}$ of $N$ nodes, whose pattern of interconnections is described by the adjacency matrix -- with generic element $A_{ij}$ giving the weight of the link from node $i$ to node $j$ (in the case of binary networks, $A_{ij}=1$ if the link $i\to j$ exists and $0$ otherwise). 
A Markov chain on a network is a discrete-time stochastic process that defines a temporal sequence of nodes (the possible states of the chain), and that satisfies the Markov property: the probability to be in any state at a given time step depends only on the state attained at the previous step. 
The process is thus described by the set of probabilities $\{p_{ij}\}_{i,j\in\mathcal{N}}$ of jumping from node $i$ to node $j$ at a given time step \footnote{We limit our analysis to time-homogeneous Markov chains, for which the transition probabilities do not depend on the current time step.}.

A Markov chain is {\em ergodic} if it is non-periodic and in the long time regime it visits each node of the network with a non-zero frequency, which converges to a stationary distribution $\{\pi_i\}_{i\in\mathcal{N}}$ satisfying the eigenvalue relation $\pi_j = \sum_{i\in\mathcal{N}}\pi_i p_{ij}$. The transition probability from node $i$ to node $j$ in a finite number $n$ of jumps, $p_{ij}^n$, is obtained from the $n$-th power of the single jump transition probability matrix. 
Similarly, the expected proportion of times that a chain starting from node $i$ visits node $j$ in the first $n$ jumps is $q_{ij}^n={n}^{-1}\sum_{m=1}^{n}p_{ij}^{m}$. Because of ergodicity, in the long-time limit both these quantities converge to the stationary frequency of visiting node $j$ (which is independent on the initial node $i$):
\begin{equation}
	\lim_{n\to\infty} p_{ij}^n=\lim_{n\to\infty}q_{ij}^n=\omega_{ij}\equiv\pi_j,
\end{equation}
where $\omega_{ij}$ is the {\em infinite time} transition probability. Note however that the convergence of $p_{ij}^n$ is exponential with $n$, whereas, that of $q_{ij}^m$ is algebraic in $m$. 
Finally, the stationary {\em probability flux} from node $i$ to node $j$ is the probability that the chain actually jumps from $i$ to $j$, and thus is given by the asymptotic joint probability of being in $i$ and successively jump to $j$:
\begin{equation}
	\mathrm{F}_p(i\to j) = \pi_ip_{ij}.
\end{equation}

\subsection*{Lumped Markov chain on network partitions}

A partition of the network nodes into a set of communities $\{\mathcal{C}\}$ induces an aggregated dynamical process, described by transition probabilities $\{\tilde{p}_{\mathcal{C}\mathcal{C}'}\}$ between communities, that is a function of the original Markov chain. Such aggregated process is not necessarily Markovian, since the new transition probabilities could in principle depend on the whole sequence of visited nodes. However, if the original Markov chain is ergodic, it is possible to define a lumped Markov process by preserving the probability fluxes between communities \cite{Piccardi2011}:
\begin{equation}
\mathrm{F}_{\tilde{p}}(\mathcal{C}\to\mathcal{C}')=\sum_{i\in\mathcal{C}}\sum_{j\in\mathcal{C}'}\mathrm{F}_p(i \to j). 
\label{eq:zeroprocess}
\end{equation}
This property is called weak lumpability, and translates into a transition probability from $\mathcal{C}$ to $\mathcal{C}'$ of the form
\begin{equation}
	\tilde{p}_{\mathcal{C}\mathcal{C}'} = \frac{\sum_{i\in\mathcal{C}}\sum_{j\in\mathcal{C}'}\pi_ip_{ij}}{\tilde{\pi}_{\mathcal{C}}},
	\label{eq:PartitionedMarkovProcess}
\end{equation}
where $\tilde{\pi}_{\mathcal{C}}=\sum_{i\in\mathcal{C}}\pi_i$. 
The generic diagonal term of this matrix, $\tilde{p}_{\mathcal{C}\mathcal{C}}$, is the {\em persistence probability} of community $\mathcal{C}$ \cite{Kim2010,Piccardi2011}. Analogously, we can build the finite and infinite time transition probabilities of the lumped process as $\tilde{p}^n_{\mathcal{C}\mathcal{C}'}={\tilde{\pi}_{\mathcal{C}}}^{-1}\sum_{i\in\mathcal{C}}\sum_{j\in\mathcal{C}'}\pi_i p^n_{ij}$, $\tilde{q}^n_{\mathcal{C}\mathcal{C}'}={\tilde{\pi}_{\mathcal{C}}}^{-1}\sum_{i\in\mathcal{C}}\sum_{j\in\mathcal{C}'}\pi_i q^n_{ij}$ and $\tilde{\omega}_{\mathcal{C}\mathcal{C}'} = {\tilde{\pi}_{\mathcal{C}}}^{-1}\sum_{i\in\mathcal{C}}\sum_{j\in\mathcal{C}'}\pi_i\omega_{ij}$ \footnote{This is generally different from taking the limit $n\to\infty$ of the $n$ jump transition probabilities for the lumped process}.

\subsection*{The natural Markov chain}

The natural Markov chain gives the simplest instance of transition probabilities between nodes in a network: the probability $p_{ij}$ to jump from node $i$ to one of its neighbor $j$ is uniform across the neighbors, and zero for unconnected nodes. If the network is weighted, $p_{ij}$ simply becomes proportional to $A_{ij}$. Hence in general $p_{ij}=A_{ij}/d_i$ where $d_i=\sum_{j\in\mathcal{N}}A_{ij}$ denotes the total weight of outgoing connections for node $i$. 

In the case of undirected networks ($A_{ij}=A_{ji}$ for each $i,j$) with a single connected component, the natural Markov chain is ergodic and the stationary distribution has the analytic form $\pi_i = d_i/(2L)$, where $2L=\sum_{j\in\mathcal{N}}d_j$. Besides, the chain is {\em reversible} since it satisfies the detailed balance: the probability fluxes between any two nodes are equal, $\mathrm{F}_p(i\to j)\equiv\pi_ip_{ij}=\pi_jp_{ji}\equiv\mathrm{F}_p(j\to i)$. 
This relation holds simply because fluxes are proportional to the elements of the adjacency matrix, $\mathrm{F}_p(i\to j)\sim A_{ij}$. 

Due to this property, the lumped process of a natural Markov chain on a network partition can be mapped to a new weighted adjacency matrix, whose terms are given by the sum of elements of the original adjacency matrix corresponding to nodes in the considered partitions:
\begin{equation}
\mathrm{F}_{\tilde{p}}(\mathcal{C}\to\mathcal{C}')\sim \tilde{A}_{\mathcal{CC}'} = \sum_{i\in\mathcal{C}}\sum_{j\in\mathcal{C}'} A_{ij}.
	\label{eq:GroupingAdjMG}
\end{equation}

\section{Community Detection with Lumped Markov chains}
\label{sec:community_detection}

We now use the general dynamical framework of lumped Markov chain introduced above to define a quality function for community detection tasks. We start from two key assumptions on which we base our definition of communities.

Firstly, as stated above, different Markov dynamics induce different partitions of the network. According to the principle behind the Markov stability, communities are regions of the network where the Markov chain remains confined for relatively long time -- where ``relatively long'' has to be assessed using a {\em reference process}. 
The most natural choice is to use a reference that brings {\em zero information} both on the details of the network topology and on the initial state of the dynamics. 
The infinite time transition probabilities of eq.~\eqref{eq:zeroprocess} satisfy this requirement, but this is just a possible choice -- we can as well use as reference the Markov dynamics at any finite time.

Secondly, we require that any community has to be resilient to changes occurring locally elsewhere in the network, or equivalently that a community is a community almost independently on the topological details of the rest of the network. 
This assumption allows simplifying the assessment of an individual community using a lumped Markov chain with two states: the community itself and the rest of the network. 
Notably such a two-states Markov process can be described using only the stationary distribution $\tilde{\pi}_{\mathcal{C}}$ and persistence probability $\tilde{p}_{\mathcal{C}\mathcal{C}}$ of the community concerned, since the conservation of probability fluxes implies that the flux from $\mathcal{C}$ to anywhere else is equal to the flux from anywhere else to $\mathcal{C}$ \footnote{An ergodic Markov chain with two possible states (labeled 1 and 2) can be described using only two parameters. 
Here we take $\pi_1$, the stationary distribution of state 1, and $p_{12}$, the transition probability from state 1 to 2. Firstly, because of normalization we have $\pi_2=1-\pi_1$ and $p_{11} = 1-p_{12}$. Then, since the chain is reversible we have $\mathrm{F}(1\to2)=\mathrm{F}(2\to1)$ and thus $p_{21}=p_{12}\pi_1/(1-\pi_1)$ and $p_{22}=1-p_{21}$.}.

\subsection*{Generalized Markov stability (GMS)}

To find a good network partition, we aim at ``maximizing the difference'' between the Markov dynamics and the reference process. 
We can thus build a quality function based on the probability flux difference between these two processes. 
For simplicity we start by considering the single jump Markov dynamics, using as reference process its asymptotic behavior given by the infinite time transition probabilities. 
For any two nodes $i,j$ in the original Markov chain we define 
\begin{equation}
	D_{ij} = \mathrm{F}_{p}(i \to j) - \mathrm{F}_{\omega}(i \to j)=\pi_i(p_{ij}-\omega_{ij}),
\end{equation}
while for two communities $\mathcal{C},\mathcal{C}'$ in the lumped chain
\begin{equation}
	D_{\mathcal{C}\mathcal{C}'} = \mathrm{F}_{\tilde{p}}(\mathcal{C}\to\mathcal{C}') - \mathrm{F}_{\tilde{\omega}}(\mathcal{C}\to\mathcal{C}')=\tilde{\pi}_{\mathcal{C}}\left(\tilde{p}_{\mathcal{C}\mathcal{C}'}-\tilde{\omega}_{\mathcal{C}\mathcal{C}'}\right)
	\label{eq:ModularityLike}
\end{equation}
meaning that $D_{\mathcal{C}\mathcal{C}'}\equiv\sum_{i\in\mathcal{C}}\sum_{j\in\mathcal{C}'}D_{ij}$. 
From this definition we see that the flux difference internal to community $\mathcal{C}$, $D_{\mathcal{C}\mathcal{C}}$, satisfies the requirement of depending only on quantities related to $\mathcal{C}$ itself -- with respect to the rest of the network. 
As global quality function to assess the quality of a network partition we can thus take the trace
\begin{equation}
    \mathcal{M}^{[1,\infty]}(\{\mathcal{C}\})=\sum_{\mathcal{C}}\tilde{\pi}_{\mathcal{C}}\left(\tilde{p}_{\mathcal{C}\mathcal{C}}-\tilde{\omega}_{\mathcal{C}\mathcal{C}}\right),
	\label{eq:ModularityLike_Trace}
\end{equation}
representing the probability flux that a random walker remains in a community within one time step, discounting the stationary distribution of the process. 
More generally, if we consider transition probabilities of $n$ jumps against visiting frequencies of $m$ jumps (with $n<m$) we can define
\begin{equation}
    \mathcal{M}^{[n,m]}(\{\mathcal{C}\})=\sum_{\mathcal{C}}\tilde{\pi}_c\left(\tilde{p}_{\mathcal{C}\mathcal{C}}^n-\tilde{p}^m_{\mathcal{C}\mathcal{C}}\right).
    \label{eq:GMS}
\end{equation}
This quality function is a {\em generalized Markov stability} (GMS). Indeed for $m\to\infty$ the reference process is given by the infinite time transition probability as in eq.~\eqref{eq:ModularityLike_Trace}, which converges to the stationary distribution of the dynamics, and in this case $\mathcal{M}^{[n,\infty]}(\{\mathcal{C}\})$ coincides with the traditional definition of Markov stability~\cite{Delvenne2010,Lambiotte2014}. 
Also, for a natural Markov chain on an undirected network and $n=1$, $\mathcal{M}^{[1,\infty]}(\{\mathcal{C}\})$ coincides with the standard modularity~\cite{Newman2003}. 
This equivalence holds because the modularity relies on a null network model (known as the {\em Chung-Lu configuration model}) that constrains node degrees~\cite{Newman2003}, and the expectations of link probabilities under this null model coincide with the infinite time transition probabilities of the natural Markov chain. 

The Markov stability and its generalized version have conceptual and practical advantages with respect to modularity. 
First of all, the modularity is based on simple link counts, as well as on a null model for the network topology. 
Typically, null model implementations are limited to the simple Erd{\"o}s-R{\'e}nyi random graph, the configuration model and the (possibly degree-corrected) stochastic blockmodel~\cite{Newman2016a} -- the few cases that have an analytic formulation. 
The Markov stability is instead based on a generic Markov process on the network: besides the natural Markov chain one is free to consider other dynamics, \eg, PageRank~\cite{Page1998} or maximal entropic random walks~\cite{Burda2009}, as well as higher-order Markov models \cite{salnikov2016HO}. 
Moreover, GMS is automatically defined in the case of directed networks, at stake with modularity~\cite{Kim2010,Lambiotte2014}.

The peculiar advantage of our generalization of Markov stability is instead the possibility of choosing the reference process, in particular by setting its time horizon. This last aspect in particular relates to the resolution limit of the modularity. 
According to~\cite{Fortunato2010}, \textit{``the resolution limit comes from the very definition of modularity, in particular from its random model. The weak point of the random model is the implicit assumption that each vertex can interact with every other vertex, which implies that each part of the network knows about everything else [...] It is certainly more reasonable to assume that each vertex has a limited horizon within the network''}. In terms of Markov stability, this is implemented as in eq.~\eqref{eq:GMS} by using a finite time horizon for the reference process. In this way, the dynamics is compared not to its stationary distribution (which is achieved after the walker has explored the whole network), but to the finite-time frequency of visiting nodes (\ie, what the walker is able to explore in a finite number of steps). Thus, it is possible to find smaller communities than with modularity. 
Note also that previous attempts~\cite{Delvenne2010,Lambiotte2014} to overcome the resolution limit with standard Markov stability are based on a continuous-time process in the small time limit, rather than on a different reference process as in eq.~\eqref{eq:GMS}.

Finally, the definition of the quality function using lumped Markov chains is invariant under hierarchical partitioning of the network. We use this feature when implementing the numerical search of communities (see pseudocode \ref{GMS_algo}) using a variant of the Louvain algorithm~\cite{Blondel2008} and a coarse-graining procedure.

\begin{algorithm}[H]
  \caption{Louvain-based algorithm}  \label{GMS_algo}
\begin{algorithmic}
\Procedure{GMS Maximization}{}
\State{\em input:}
\State $\{p\} \gets \text{transition probabilities between nodes}$
\State $\{\mathcal{C}\}$: initial partition of single-node communities
\State $list$: community membership of each node
\Repeat
    \State $\{\tilde{\mathcal{C}}\} \gets $ \Call{Moves}{$\{\mathcal{C}\},\{p\}$}
    \If {$\mathcal{M}(\{\tilde{\mathcal{C}}\},\{p\})>\mathcal{M}(\{\mathcal{C}\},\{p\})$}
        \State update $list$ according to $\{\tilde{\mathcal{C}}\}$
        \State $\{p\} \gets \text{lumped transition probabilities -- eq. \eqref{eq:PartitionedMarkovProcess}}$
        \State $\{\mathcal{C}\} \gets \text{coarse-grained $\{\tilde{\mathcal{C}}\}$ (one node per community)}$
        \EndIf
    \Until{$\mathcal{M}$ reaches a maximum}
\State{\bf output} $list$
\State{\em final step:}
\State $\{p\} \gets \text{transition probabilities between nodes}$
\State $\{\mathcal{C}\}$: partition corresponding to $list$
\State $\{\mathcal{C}\} \gets $ \Call{Moves}{$\{\mathcal{C}\},\{p\}$}  
\EndProcedure
\State
\Function{Moves}{$\{\mathcal{C}\},\{p\}$}
\State{(repeat a few times)}
\ForAll{communities $\mathcal{C}\in\{\mathcal{C}\}$}
    \ForAll{nodes $i\in \mathcal{C}$}
        find $\mathcal{C}'\neq\mathcal{C}$ {\bf such that}
        \State moving $i$ from $\mathcal{C}$ to $\mathcal{C}'$ maximally increases $\mathcal{M}$
        \If{such $\mathcal{C}'$ exists} move $i$ from $\mathcal{C}$ to $\mathcal{C}'$
            \EndIf
        \EndFor
    \EndFor
\EndFunction
\end{algorithmic}
\end{algorithm}

\subsubsection*{Numerical optimization}\label{subsec:Algorithm}

We first work at the finest level of nodes. We start with a configuration where each node is considered as a different community, giving the corresponding initial value for $\mathcal{M}^{[n,m]}$. 
The moves we consider are successive changes of community for individual nodes. Each move is accepted if the induced change to $\mathcal{M}^{[n,m]}$ is positive (such variation is computed locally because we consider only moves of single nodes and not of node groups). These moves are repeated until no further increase of $\mathcal{M}^{[n,m]}$ can be achieved.

The communities found through this first procedure are then taken as the meta-nodes of a coarse-grained network, while the Markov process for this new network is defined using the lumpability condition of eq.~\eqref{eq:PartitionedMarkovProcess}. The local moves described above are then repeated again for this network until a new maximum of $\mathcal{M}^{[n,m]}$ is reached. The corresponding partition is then used to build a more coarse-grained network, and the whole process is repeated until no further moves nor coarse-graining steps can increase $\mathcal{M}^{[n,m]}$.

As a final step, we restart the method from the node level but imposing the community structure just found -- that is, we check whether the move of a single node can refine the optimal partition. This is for instance the case in the Karate Club network \cite{konect:ucidata-zachary} (see below), in which a single node switches community because of this last step and the GMS value rises from $0.4188$ to $0.4198$.

\section{Results}
\label{sec:Results}

\subsection*{Resolution of generalized Markov stability}
\label{subsec:resolution}

\begin{figure}[t!]
	\centering
	\includegraphics[width=0.36\textwidth]{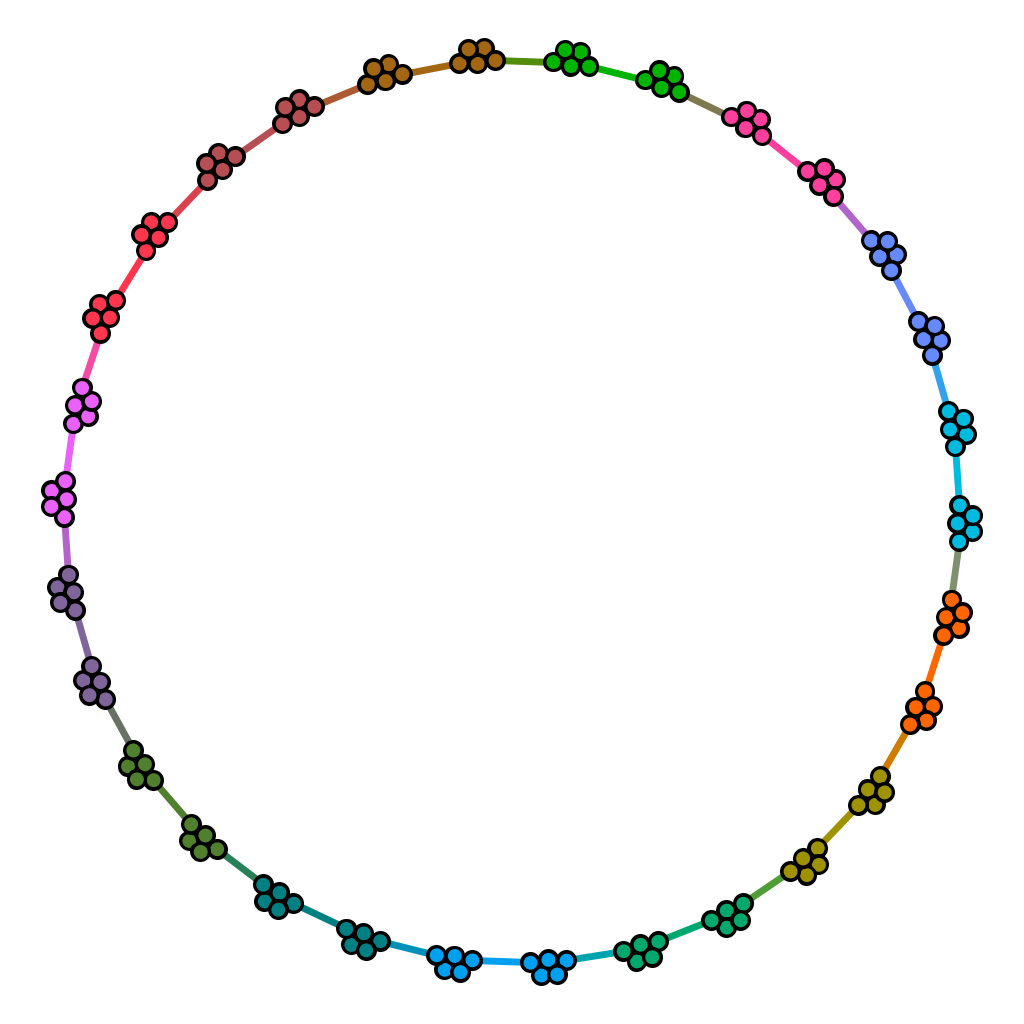}\\
	\vspace{5mm}
	\includegraphics[width=0.5\textwidth]{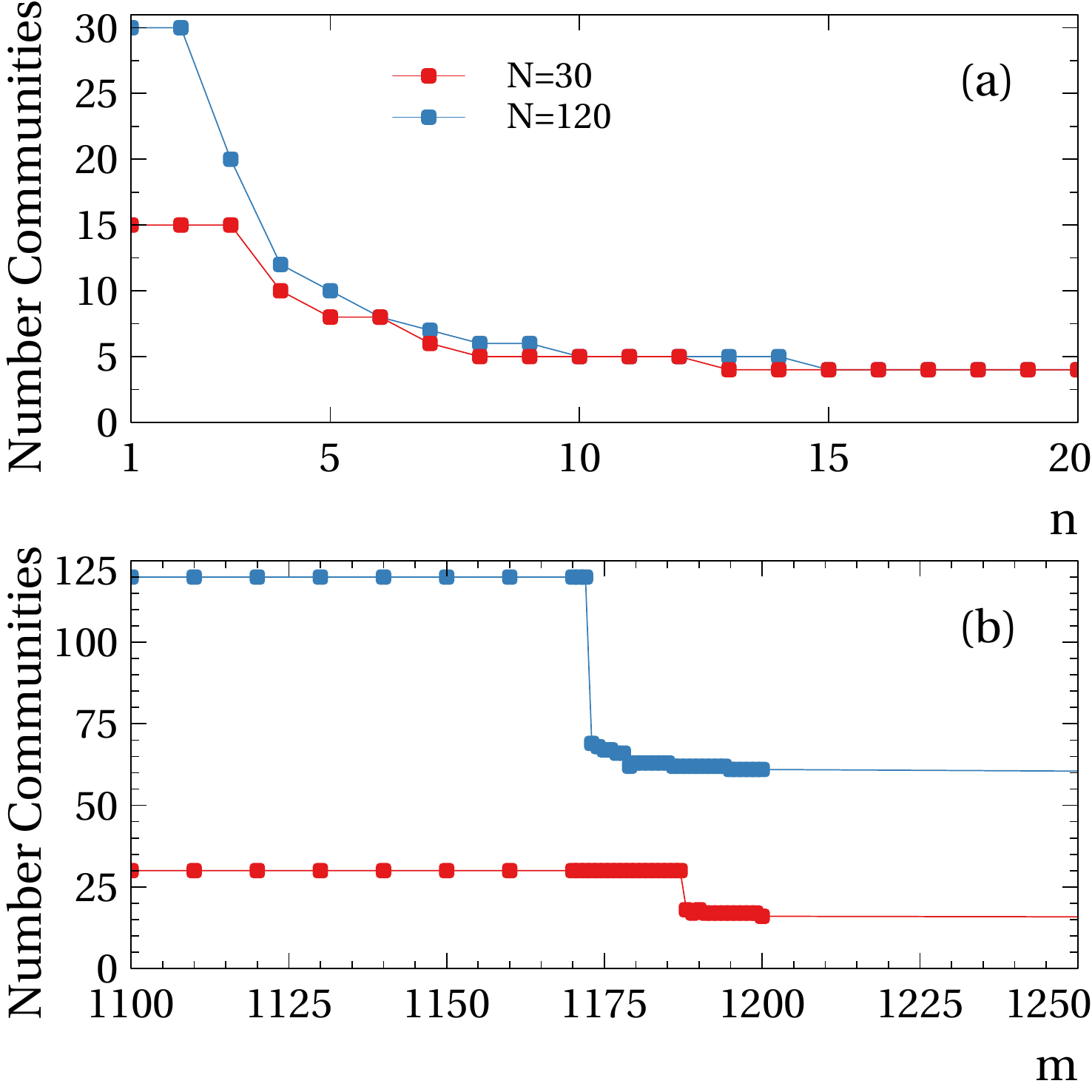}
	\caption{Resolution of generalized Markov stability for a natural Markov chain on the illustrated network (a ring-like configuration with $N=30$ cliques of $5$ nodes, each clique being connected to only two other cliques). 
	Panel $(a)$: number of communities found by $\mathcal{M}^{[n,\infty]}$ as a function of $n$.
	Panel $(b)$: number of communities found by $\mathcal{M}^{[1,m]}$ as a function of $m$. 
	In both panels we show the cases $N=30$ (red lines) and $N=120$ (blue lines).}
	\label{fig:Resolution_Jumps}
\end{figure}

We first explore the resolution of $\mathcal{M}^{[n,m]}$ with respect to different choices of Markov times $n$ and $m$. 
To this end, we consider a natural Markov chain process on the standard toy network of maximal modularity~\cite{Fortunato2007}: a ring-like configuration with $N$ cliques of $5$ nodes, each clique being connected to only two other cliques (Figure~\ref{fig:Resolution_Jumps}). 
For this graph with $N=30$ cliques, standard modularity optimization returns a structure of $15$ communities (each composed by a pair of cliques), whereas, for $N=120$ modularity returns $30$ communities (each aggregating four adjacent cliques). 
Standard Markov stability $\mathcal{M}^{[n,\infty]}$ for $n>1$ instead finds a community structure that is coarser than what is found by modularity \cite{Delvenne2010}. In particular, since communities are defined as regions of the network where the walker remains confined within $n$ jumps, communities become less in number and bigger in size by increasing the time horizon $n$ of the dynamics -- as shown in panel (a) of Figure~\ref{fig:Resolution_Jumps}. 
Notably, if we use a reference process at finite time, we automatically obtain finer communities than with modularity. 
Panel (b) of Figure~\ref{fig:Resolution_Jumps} shows the case of the simple function $\mathcal{M}^{[1,m]}$: there is a sharp transition for the number of communities at a critical value $m^*$, below which the true structure of $N$ communities (one for each clique) emerges.

\begin{figure*}[t!]
	\centering
	\large 
	(a)
	\includegraphics[width=0.35\textwidth]{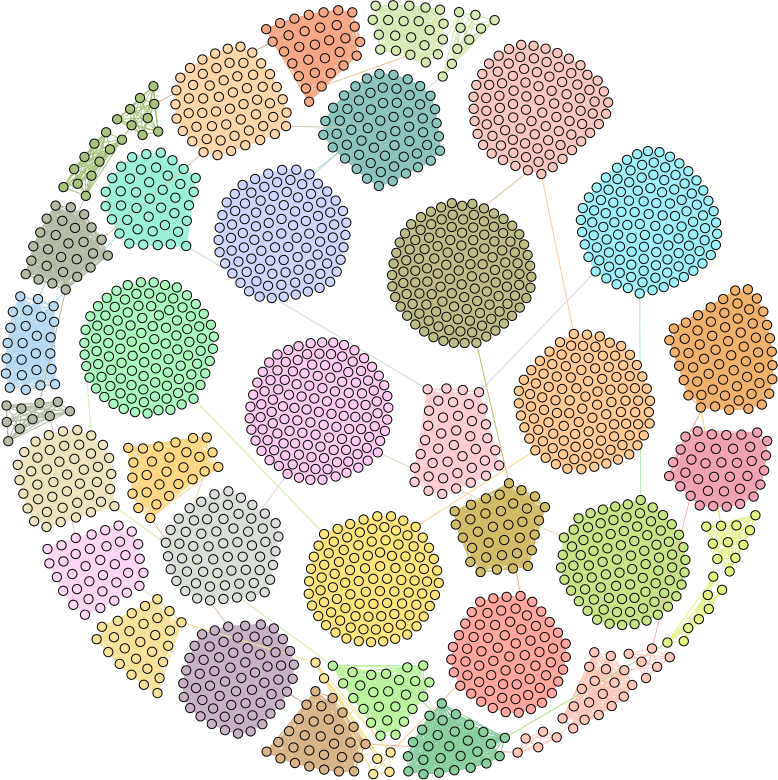}
	\hspace{2cm}
	(b)
	\includegraphics[width=0.35\textwidth]{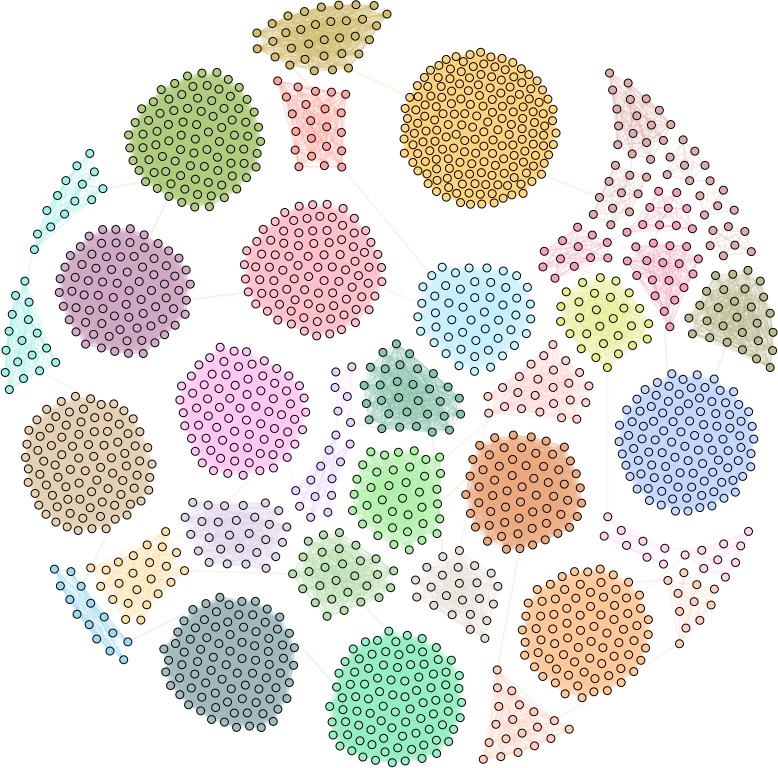}
	\\
	\vspace{5mm}
	\includegraphics[width=\textwidth]{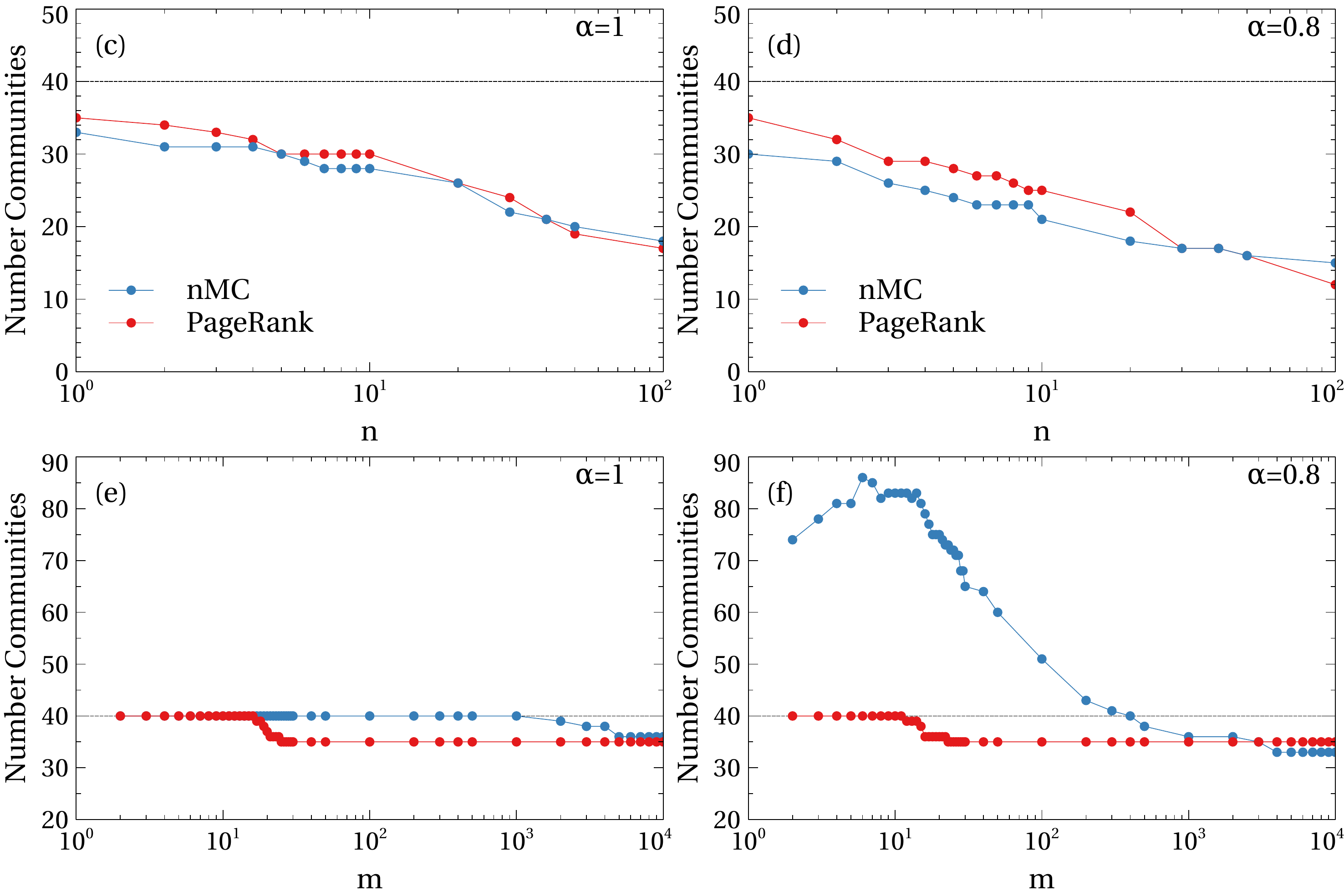}
	\caption{(a,b) Visual representation of communities found by standard modularity $\mathcal{M}^{[1,\infty]}$ on a ring-like configuration with $40$ cliques of varying (exponentially distributed) size, each clique being connected to only two other cliques. The internal connection probability $\alpha$ of the cliques is 1.0 for network (a) and 0.8 for network (b). Each community is represented by a different color. 
    (c,d,e,f) Number of communities identified by generalized Markov stability $\mathcal{M}^{[n,m]}$ as a function of parameters $n$ and $m$ (the time scale of the dynamics and of the reference process) for the same network configuration of panels (a,b). 
	We show GMS implementations using the natural Markov chain (in blue) and PageRank with $\mu/N=0.15$ (in red). See below for further details on this alternative dynamics.}
    \label{fig:exponential_clique_graphs}
\end{figure*}

The cliques-graphs considered above is a very simple example, especially because the cliques have the same size. 
We thus consider an heterogeneous graph of $40$ cliques whose size is exponentially distributed (ranging from 5 to 100 nodes). Standard modularity maximization on a realization of this graph returns $33$ communities, since it tends to group together the small nearest cliques. Instead $\mathcal{M}^{[1,m]}$ finds the true community structure as soon as $m\lesssim 10^3$.
Figure~\ref{fig:exponential_clique_graphs} shows a detailed analysis of this configuration. We consider cliques with internal connection probability $\alpha=1$ on the left and $\alpha=0.8$ on the right. Again the number of detected communities varies as expected with the parameters $n$ and $m$ that set the resolution of GMS. As in the previous example, by increasing $n$ while keeping $m$ fixed the method finds less communities (it fails to detect the small ones), hence we confirm that $n$ effectively sets the minimum size of detected communities. A more interesting picture is obtained by fixing $n$ while varying $m$. For $\alpha=1$ we see the same behavior as that of Figure \ref{fig:Resolution_Jumps}: the method finds the correct number of communities for finite $m$, whereas, it starts aggregating the smallest cliques when the time horizon of the reference process becomes much larger than the size of the largest clique. Instead for $\alpha=0.8$ cliques are not so strongly connected, and fluctuations may induce dense regions internal to cliques. Therefore, we observe a crossover between the region where $m$ is too small to accommodate for the larger cliques (so that the number of detected communities grows with $m$) and again the regime where $m$ is large enough and the number of detected communities decreases. These examples teach us that we cannot expect to achieve the best performance at infinite $m$, because in this case the method will discard important local information on the network, neither in general at small $m$ for which the horizon of the random walker is simply too limited.

\subsection*{GMS for different random processes}
\label{subsec:different_rw}

$\mathcal{M}^{[n,m]}$ of eq.~\eqref{eq:ModularityLike_Trace} is defined for a generic Markov process on the network -- the only requirement being the existence of the stationary distribution and of its finite-time version. The induced community structure can thus strongly depend on which process is implemented. Beyond the natural Markov chain, we considered two other processes. 

The first one is {\em PageRank}~\cite{Page1998} (see also \cite{Kim2010,Lambiotte2012}), which complements the natural Markov chain with a teleportation term allowing for jumps between any two nodes: $p_{ij}=(1-\mu)A_{ij}/d_i+\mu/N$. 
In general, teleportation increases the probability to jump outside a community, hence the number of identified communities decreases with the teleportation rate $\mu$. Additionally, PageRank leads to similar results to that if a natural Markov chain with longer time horizon, and at the same time is less sensitive to topological fluctuations (see Figure \ref{fig:exponential_clique_graphs} e,f). 
Notably, the teleportation rate $\mu$ makes the chain ergodic even if the network has disconnected components or if it is directed and has transient parts (that the walker cannot access after leaving them). 

The second Markov process we consider is the {\em maximal entropy random walk} (MERW)~\cite{Burda2009,Ochab2013merw}, also known as the Ruelle-Bowen process in discrete time \cite{Lambiotte2014}.
MERW transition probabilities are such that all trajectories of given length and given endpoints are equiprobable, and take the form $p_{ij}=(A_{ij}/\lambda)/(\psi_j/\psi_i)$ -- where $\lambda$ is the largest eigenvalue of the adjacency matrix and $\psi_i$ is the $i$-th component of the normalized eigenvector associated to $\lambda$. 
MERW has strong localization property, imprisoning the walkers in entropic wells \cite{Burda2009}. 

\begin{figure}[t!]
\large 
(a)\\
	\includegraphics[scale=0.35]{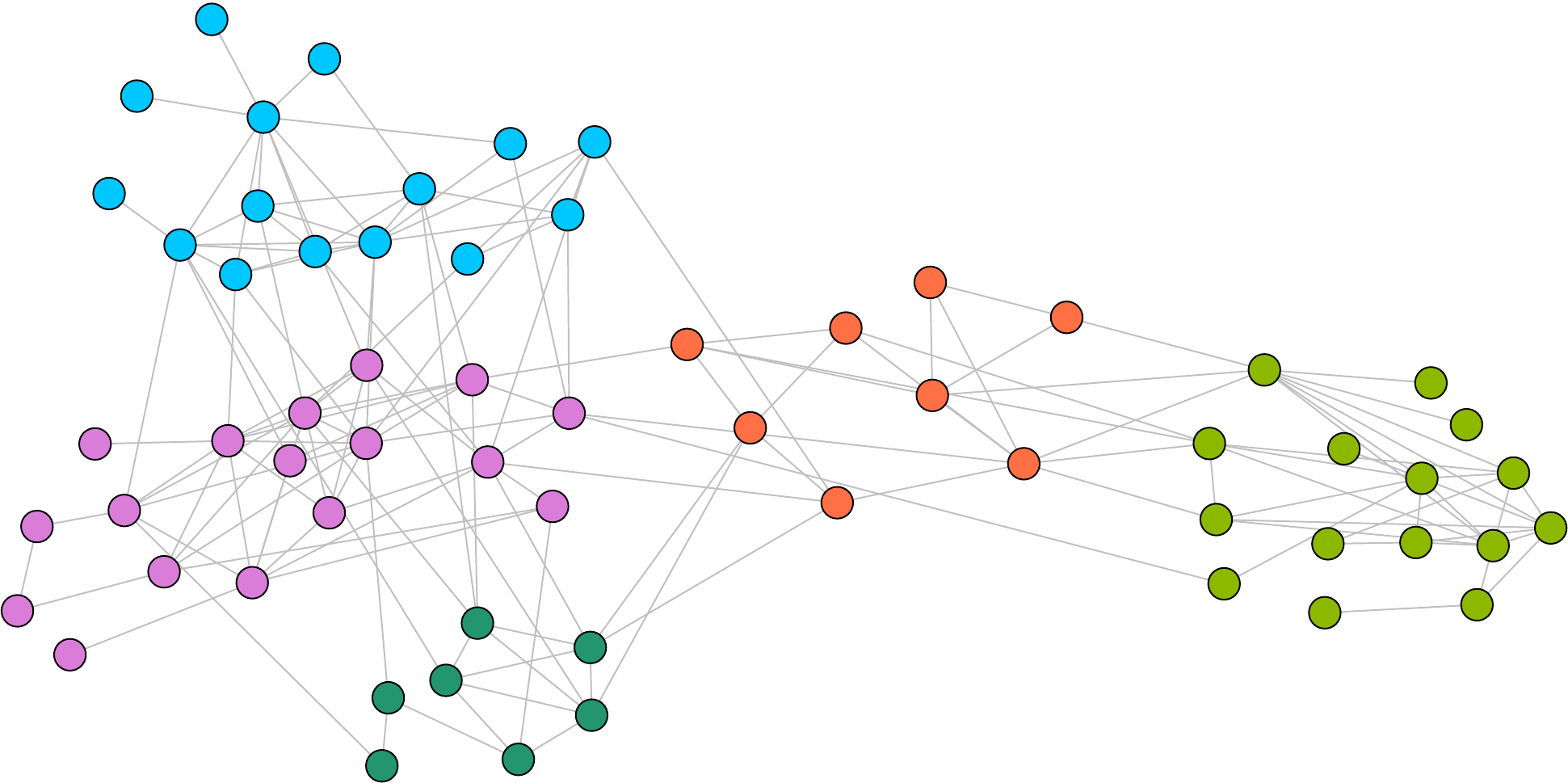}
	\vspace{0.25cm}
\\(b)\\
	\includegraphics[scale=0.35]{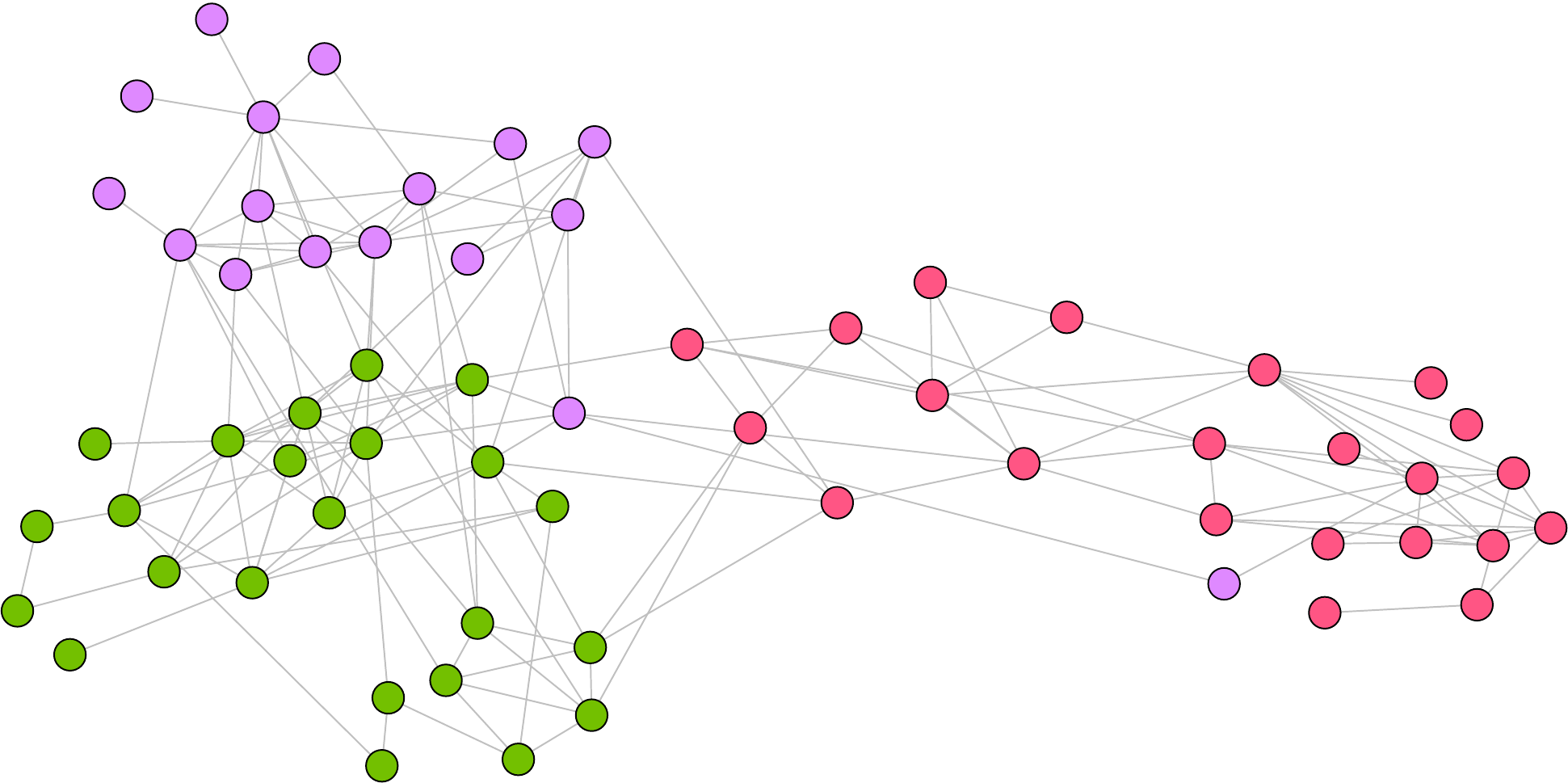} \vspace{0.25cm}
\\(c)\\
	\includegraphics[scale=0.35]{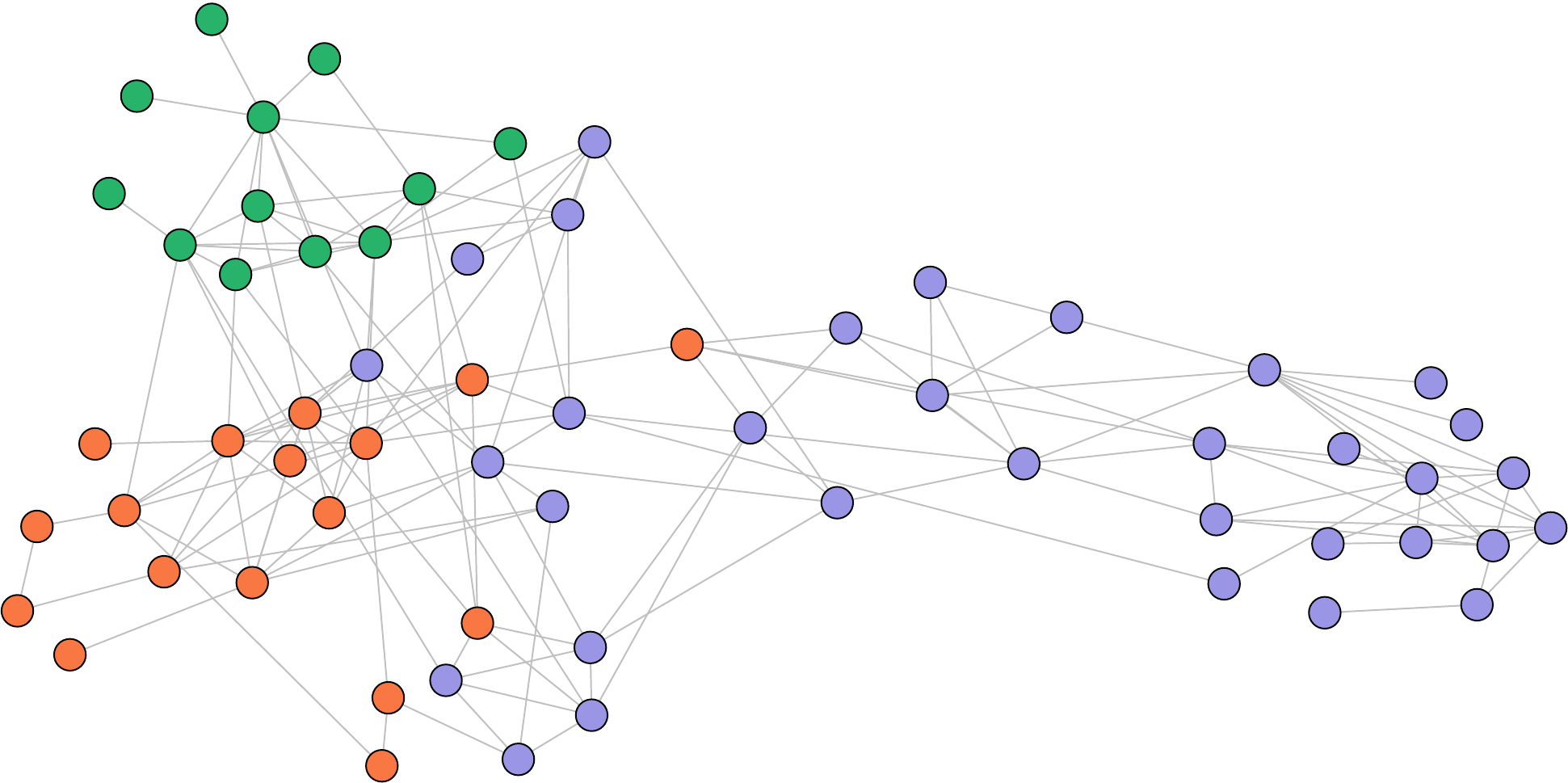}
	\caption{Communities of the \texttt{Dolphins} network found by $\mathcal{M}^{[1,\infty]}$ with natural Markov chain (a), PageRank with $\mu/N=0.006$ (b) and MERW (c).}
	\label{fig:dolphins_different_rw}
\end{figure}

To visually grasp the effect of using a particular Markov dynamics, we show in Figure~\ref{fig:dolphins_different_rw} the communities detected by $\mathcal{M}^{[1,\infty]}$ using natural Markov chain, PageRank and MERW, 
on the illustrative example of the \texttt{Dolphins} network~\cite{konect:dolphins} -- the network of ``swimming together'' relations among a group of dolphins. 
For this network we do not have information on any reference community structure, hence we cannot assess which Markov dynamics performs best. However, despite the identified partitions vary with the Markov process, notably some communities seems more persistent with respect to the specific dynamics employed (in this case, the top left part of the network). Thus comparing the results of multiple dynamics can increase our confidence level on the detected network partition.

\subsection*{GMS versus metadata partitions in real networks}
\label{subsec:real_data}

\begin{figure*}[t!]
	\centering
	\includegraphics[scale=0.75]{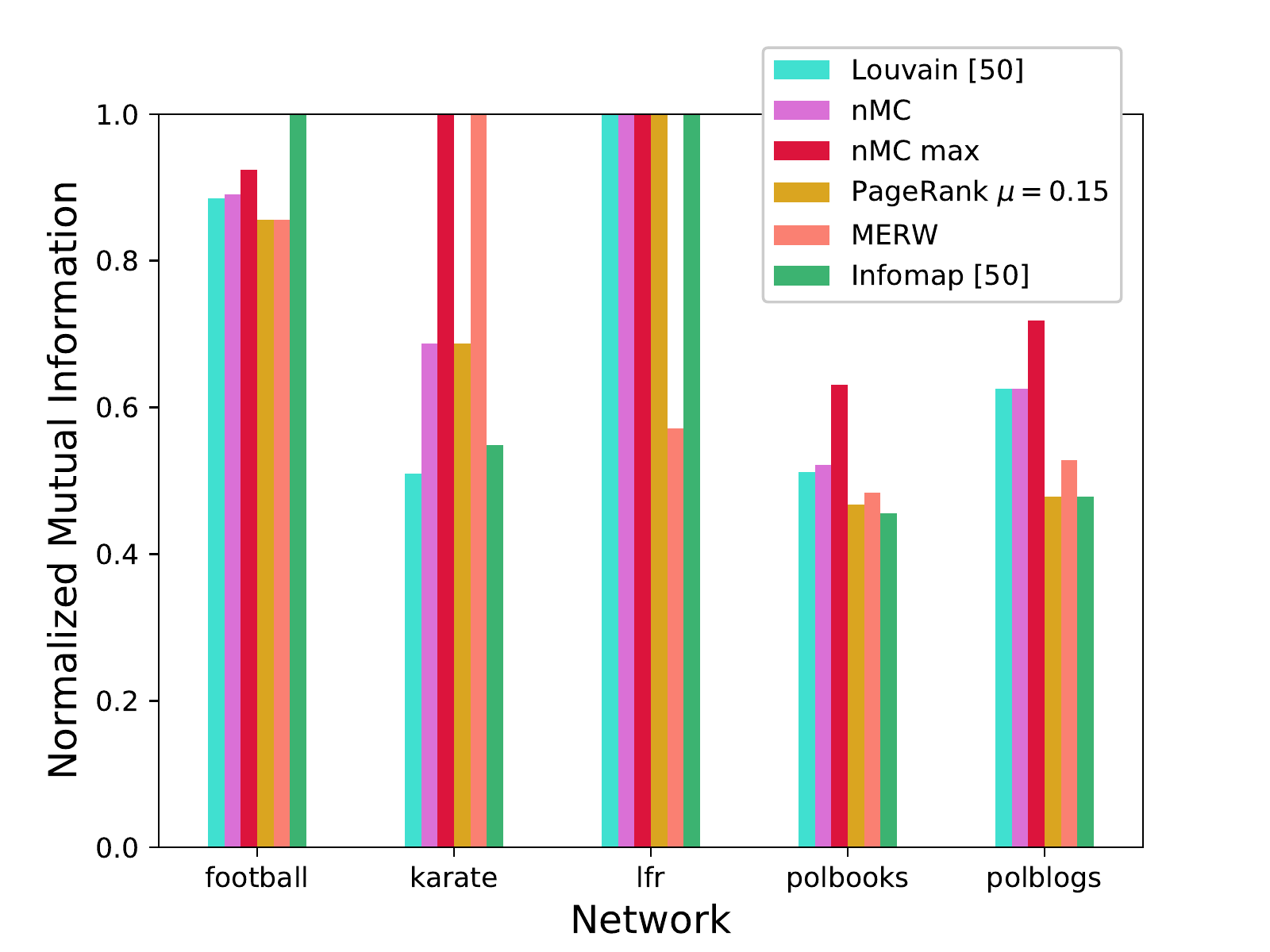}
	\caption{NMI scores between structural communities found by various methods and metadata groups, for the five networks we consider (scores are clustered by datasets on the horizontal axis). The figure is realized following Figure 5 of \cite{Hric2014}. 
	The methods used are as follows: Louvain Modularity; GMS with natural Markov chain $n=1$ $m=\infty$ (nMC); GMS with natural Markov chain and parameters $n$ and $m$ yielding the highest NMI for that network (nMC max); GMS with PageRank ($\mu/N=0.15$) $n=1$ $m=\infty$; GMS with MERW $n=1$ $m=\infty$; Infomap.}
	\label{fig:bar_plot}
\end{figure*}

We now put GMS to the test of real and synthetic networks that represent the traditional benchmarks for community detection methods. These networks posses node metadata information that allows defining reference partitions to be used for comparison (see however \cite{Hric2014,Peel2017} about the problems of associating metadata groups with topological communities). 

We start by briefly describing the datasets we use (that we downloaded from \url{http://www-personal.umich.edu/~mejn/netdata/}).  A full description can be found in the cited references, as well as in \cite{Hric2014}.
\begin{itemize}
 \item \texttt{football} is the network of American football games between Division IA colleges during season Fall 2000~\cite{Girvan2002}. Links exist if two teams played any game, and there are 12 groups of teams (conferences) for scheduling intra-group games.
 \item \texttt{karate} is the friendship network of Zachary's karate club~\cite{konect:ucidata-zachary} that has two natural communities,  corresponding to the split of the club in two factions after a dispute between the coach and the treasurer.
 \item \texttt{polblogs} is the network of (undirected) hyperlinks between weblogs on US politics after the 2004 elections~\cite{Adamic:2005:PBU}. Groups are ``liberal'' or ``conservative'' as assigned by either blog directories or self-evaluation.  
 \item \texttt{polbooks} is the network of books about US politics from 2004 election, taken from Amazon.com~\cite{krebs06}.  Links represent co-purchasing of books. Groups are based on
political alignment: ``liberal'', ``neutral'', or ``conservative'', according to human evaluation.
 \item Finally, \texttt{lfr} is an artificial network with built-in topological communities, generated through the state-of-the-art LFR benchmark~\cite{LFR} (parameters $N=1000$, 40 small communities 0f size ranging between 10 and 50, and mixing parameter $\tfrac{1}{2}$). The \texttt{lfr} generator code is available at  \url{https://sites.google.com/view/santofortunato/software}.
\end{itemize}

We use the Normalized Mutual Information (NMI)~\cite{Danon_2005} to measure the similarity between the network partition induced by a community detection method and the metadata communities of the network. 
A comparative assessment of how well different methods perform according to this metric is reported in Figure \ref{fig:bar_plot}. In particular we show the resulting NMI obtained by implementing four different dynamics on GMS:  standard natural Markov chain ($n=1$, $m=\infty$, labeled nMC); natural Markov chain with parameters $n$ and $m$ yielding the highest NMI for that network (nMC max); standard PageRank with $\mu/N=0.15$ ($n=1$, $m=\infty$); MERW ($n=1$, $m=\infty$). 
We add to the comparison the two state-of-the-art Louvain~\cite{Blondel2008} and Infomap~\cite{Rosvall2008} algorithms (for the performance of other methods, we remand the reader to Figure 5 of \cite{Hric2014}). 

\begin{figure*}[p!]
	\centering
    \includegraphics[scale=0.5]{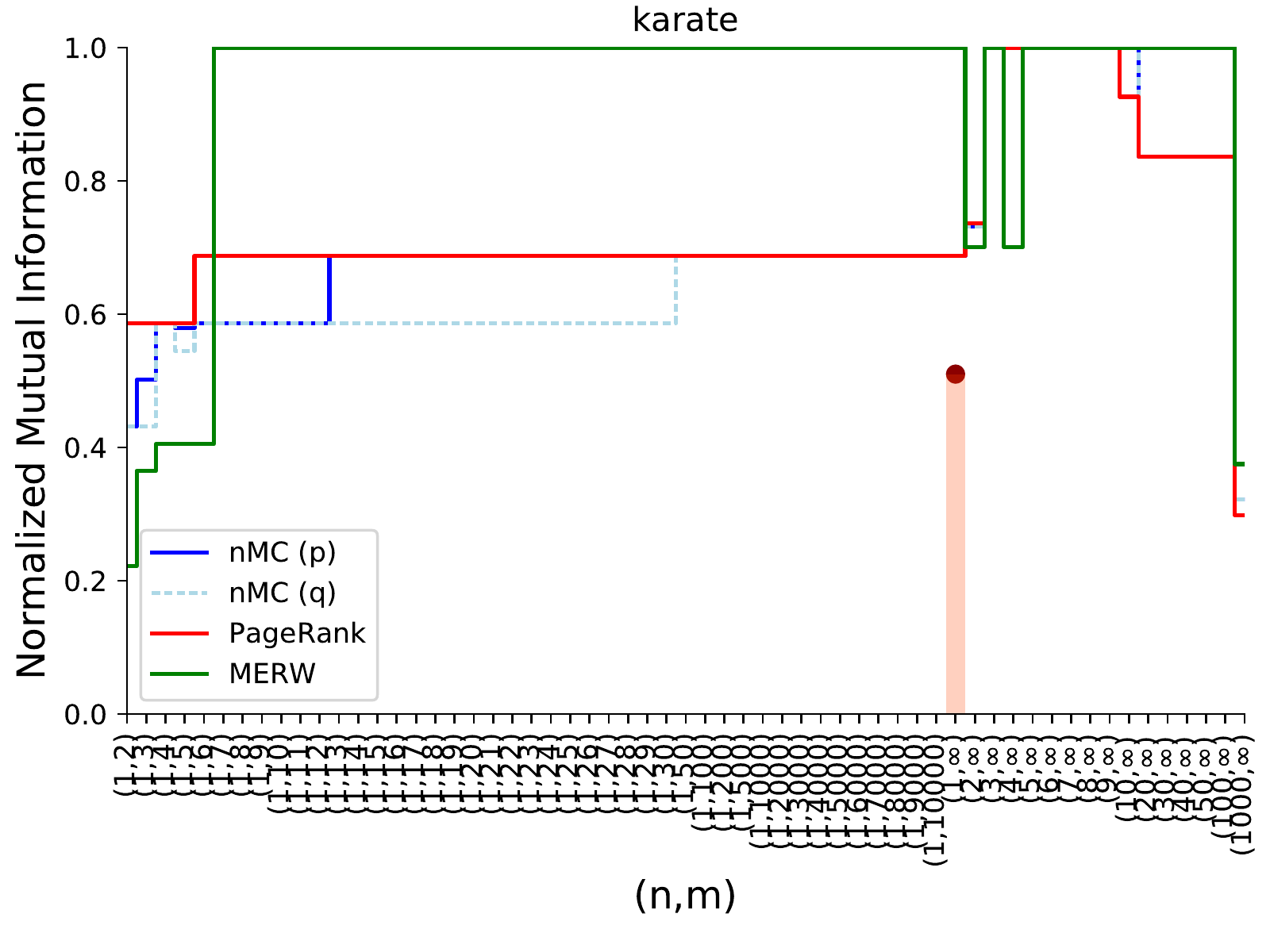}\\
	\includegraphics[scale=0.5]{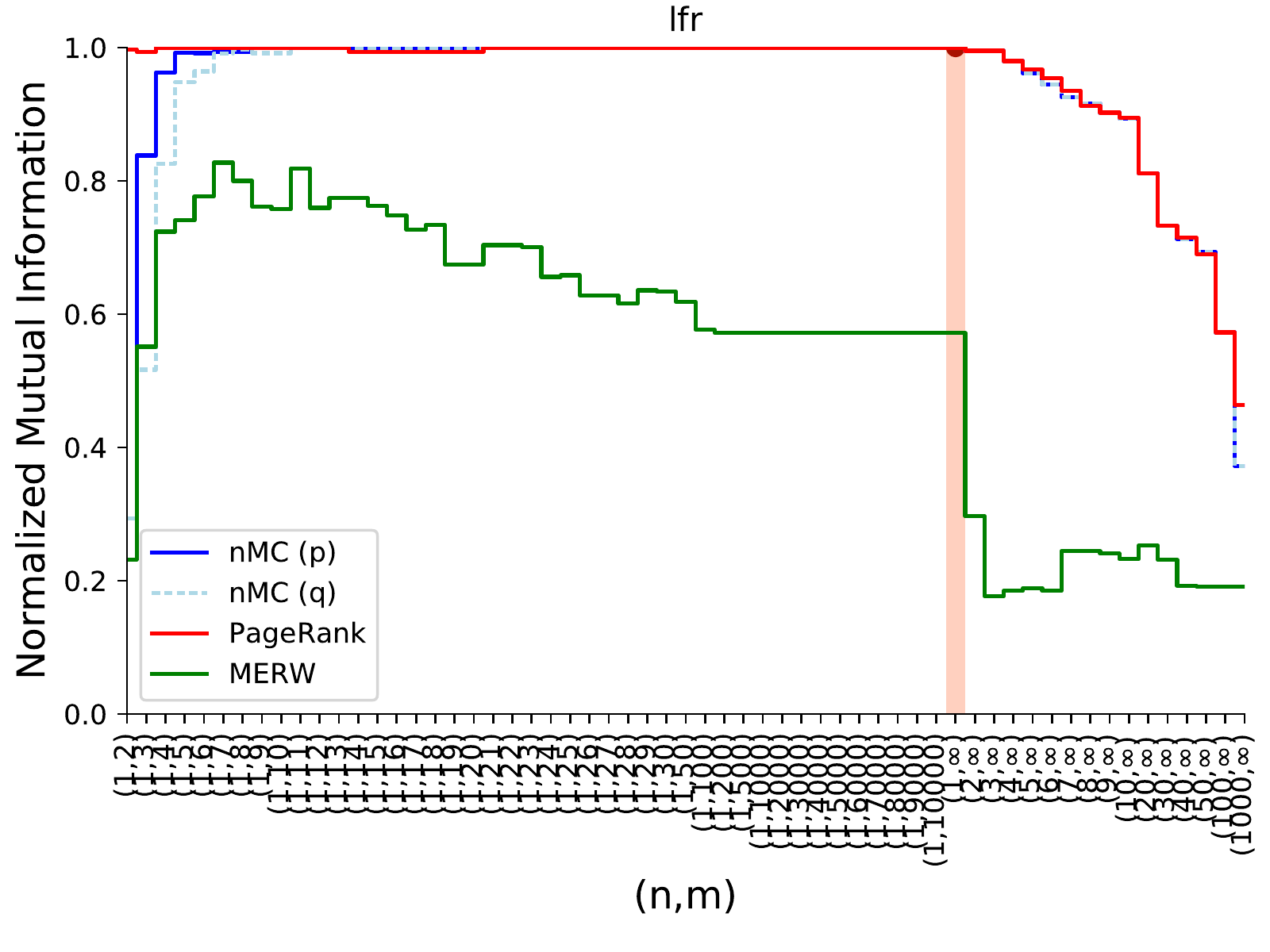}
	\includegraphics[scale=0.5]{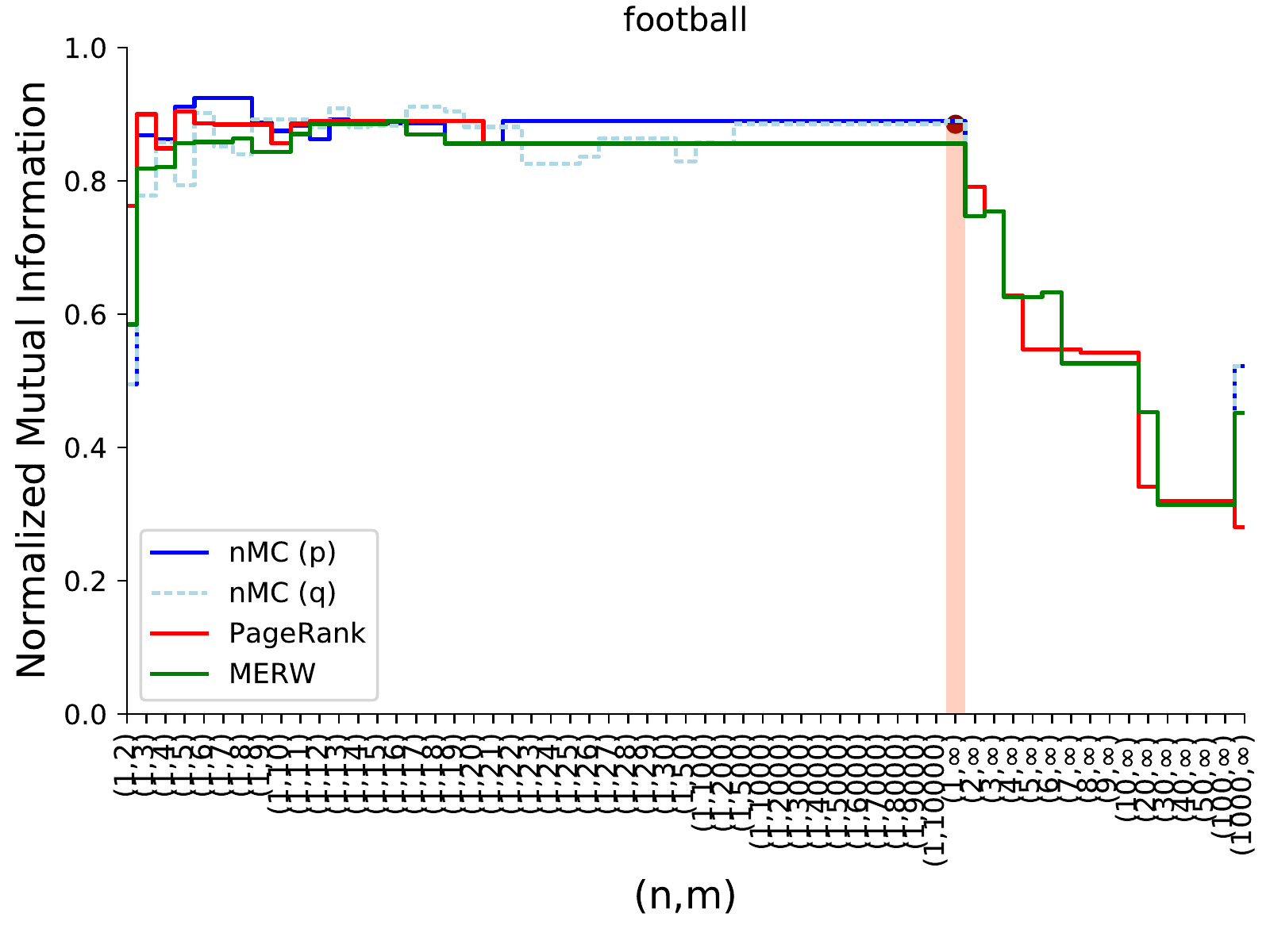}\\
    \includegraphics[scale=0.5]{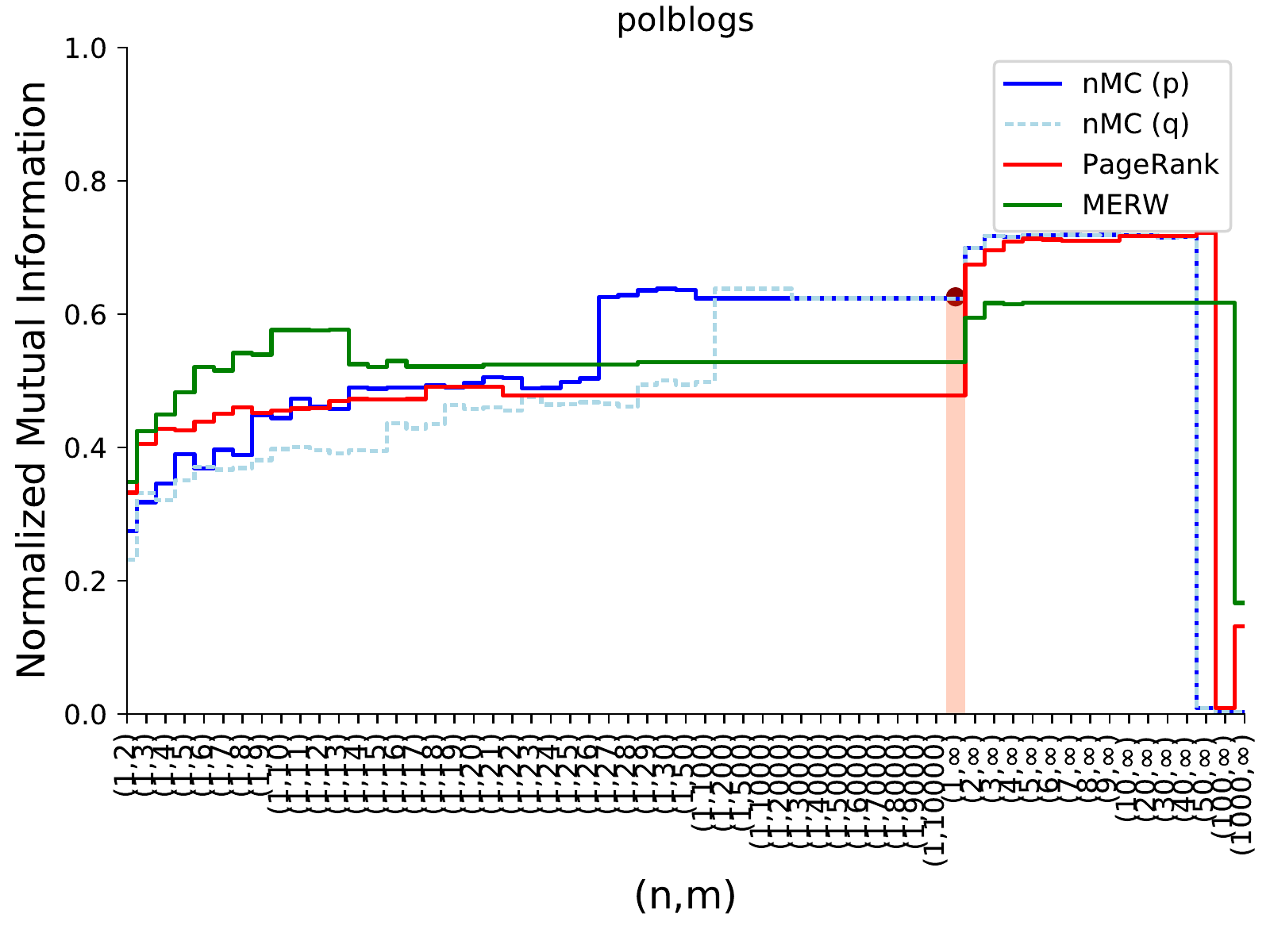}
    \includegraphics[scale=0.5]{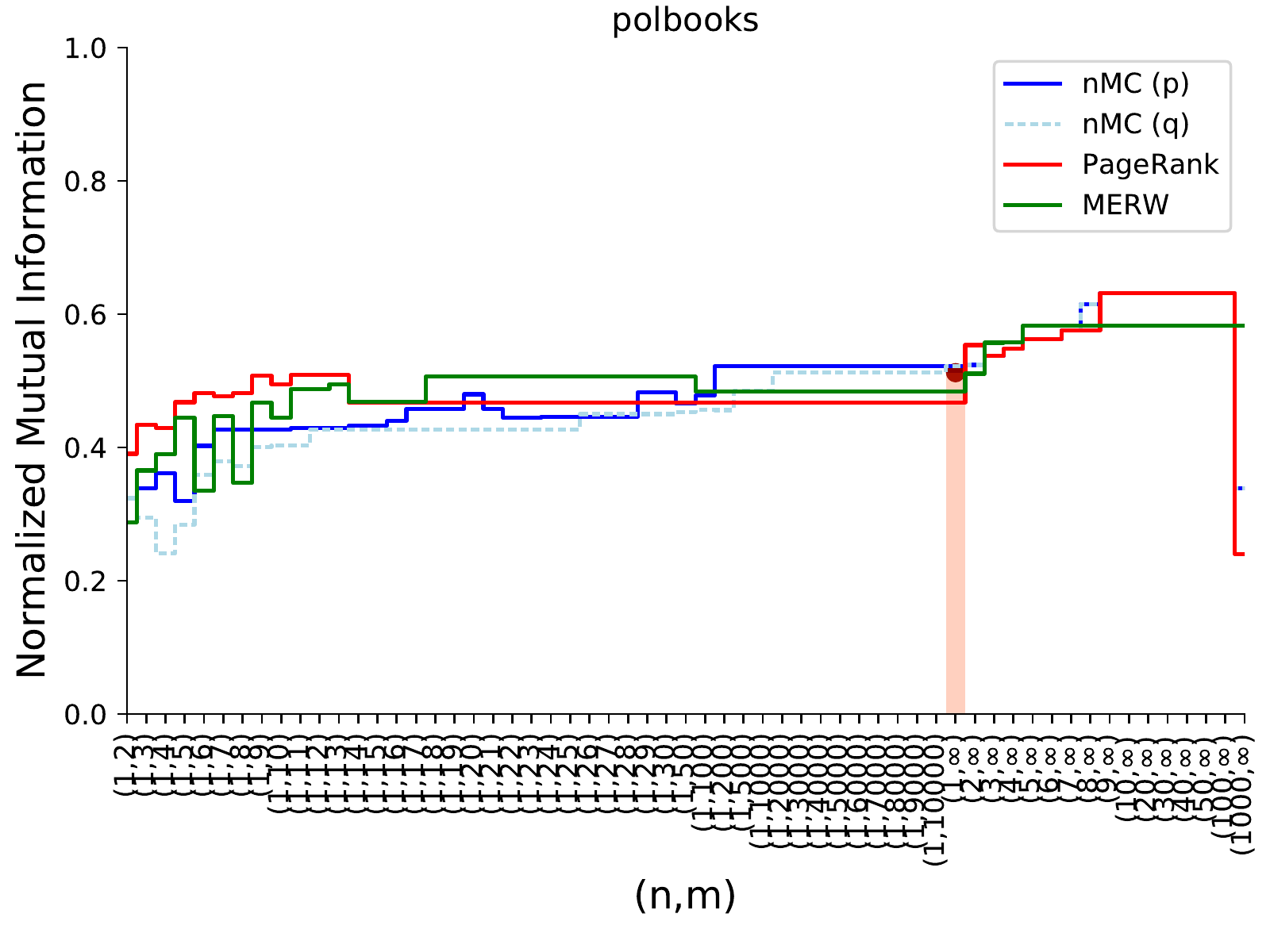}\\
	\caption{NMI scores between structural communities found by GMS and metadata communities as a function of the parameters $n$ and $m$ setting the time scale of the dynamics and of the reference process. 
	The dynamics are: natural Markov chain of eq.~\eqref{eq:GMS} (nMC (p)), natural Markov chain of eq.~\eqref{eq:GMSq} (nMC (q)), PageRank with $\mu/N=0.006$ and MERW. The red dot in correspondence of the vertical bar denotes the NMI value obtained by standard modularity using the Louvain algorithm.}
	\label{fig:varying_nm}
\end{figure*}

The detailed performance of GMS for varying $n$ and $m$ in shown in Figure \ref{fig:varying_nm} separately for each of the considered networks. 
In the case of \texttt{karate}, GMS needs $n>1$ (but finite) and $m=\infty$ to retrieve a partition corresponding to the two metadata groups. 
This is an expected outcome because the network is sparse and the two groups are big (compared to the whole network), therefore the random walker cannot fully explore them within just a few steps $n$. At the same time, $m$ must be large because each community needs to be assessed against the whole network. 
As side remark, MERW outperforms the other dynamics on most time scales. This happens because MERW is strongly localized on hubs, which in the \texttt{karate} network are the coach and the treasurer who are the central members of each group.
Moving further, in \texttt{lfr} we see on one hand that nMC and PageRank return the metadata groups of the network even at small reference horizon $m$, because each group is dense but small compared to the network size, and thus does not need to be assessed against the whole network to be retrieved. On the other hand, by increasing $n$ the NMI decreases because the random walker is more likely to travel within groups and thus GMS ends up aggregating the smallest communities. 
A similar behavior is observed in \texttt{football}: a good quality of the partition is obtained at small horizons $m$, whereas, by increasing $n$ the walker is less likely to stay confined in a group. 
Finally, at stake with the previous two networks, the \texttt{polblogs} and \texttt{polbooks} are sparse and their metadata groups are few (two for \texttt{polblogs} and three for \texttt{polbooks}) but large. Therefore, similarly to \texttt{karate}, these groups are better retrieved at moderately large values of $n$ for which the walker can fully explore each community, and at the same the reference walker must have explored the whole network ($m$ large).  

Summarizing, we have found that there is no recipe that performs best in all situations. The optimal performance of GMS as a function of the parameters $n$ and $m$ however provides information on what are the features of the communities that exist in a network. For instance, \texttt{lfr} and \texttt{football} are both characterized by many small and dense metadata groups, hence GMS work well with small scales of the dynamics ($n$) and of the reference process ($m$). \texttt{polblogs} and \texttt{polbooks} on the contrary have a few large and sparse groups, which are retrieved with wider dynamical horizon $n$. \texttt{karate} belongs to this latter case, but the presence of the two hubs and the consequent degree heterogeneity enhance the performance of MERW.

\subsection*{Alternative reference process}

GMS can be as well defined with a reference process measuring the visiting frequencies within $m$ jumps (that is, the $q^m$ matrix) instead of setting a fixed horizon at a temporal scale $m$ (represented by $p^m$). 
This leads to a reformulation of eq.~\eqref{eq:GMS} as 
\begin{equation}
    \mathcal{M}^{[n,m]}(\{\mathcal{C}\})=\sum_{\mathcal{C}}\tilde{\pi}_c\left(\tilde{p}_{\mathcal{C}\mathcal{C}}^n-\tilde{q}^m_{\mathcal{C}\mathcal{C}}\right).
    \label{eq:GMSq}
\end{equation}
In this alternative formulation, the reference process contains the contribution of short walks that carry information on the local properties of the network. As explained above, using $q$ instead of $p$ leads to a slower convergence for $m\to\infty$, however the two approaches are qualitatively similar -- especially for small values of $m$ (see Figure \ref{fig:varying_nm}).

\subsection*{Directed networks}
\label{subsec:directed_graphs}

As a final remark, we stress that the definition of generalized Markov stability does not depend on the specific network features. Therefore, GMS can be directly implemented on directed networks, provided the considered Markov chain is ergodic (an easy solution for this is the teleportation term of PageRank). 
Indeed, the case $n=1$ and $m=\infty$ for simple random walks on directed networks has been studied in~\cite{Kim2010} as a generalization of standard modularity.

\section*{Conclusions}
\label{Conclusions}

In this work we reformulated the use of ergodic Markov chains applied to the problem of community detection in networks. 
Specifically, we defined a lumped Markov process between communities, whose transition probability fluxes are built by aggregating the probability fluxes at the level of nodes. This aggregated process is then used to define a quality function to evaluate a network partition, by requiring the probability fluxes internal to communities (\ie, the persistence probabilities) to be maximally larger than those of a reference case. This results in a generalized version of the Markov stability (GMS). 

We remark that the whole theoretical construction of GMS derives from two simple requests: 1) the existence of the reference process, used to assess the persistence probabilities of the dynamics, and 2) the resilience of communities to changes occurring elsewhere in the network, so that the search of communities can be decomposed into multiple two-states problems (for each community, the assessment of the community itself against the rest of the network).

GMS can be implemented with any ergodic Markov dynamics on the network. Additionally, being based on the concept of lumped Markov chains, the GMS quality function is invariant under network partitioning. This means that we can aggregate and disaggregate both nodes and node groups without losing information on the structure and dynamics of the network. This feature is at the basis of the algorithm we developed to optimize the quality function.

Concerning the implementation of GMS, when the reference process corresponds to a dynamics in which all the information on initial conditions and nodes correlations is lost, as in the case of the infinite time transition probability, we obtain the standard formulation of the Markov stability. 
However considering a reference process with a finite time horizon allows finding communities of varying size -- thus overcoming in a natural way the resolution limit typical of the modularity and other approaches. 
Indeed the time scales of the Markov dynamics and of the reference process effectively set the resolution level of the method. Communities obtained at different resolutions are in general not hierarchical, as in \cite{Arenas2008}. However, optimizing the GMS quality function with respect to $n$ and $m$ means identifying the size window and other features of the network communities. For a given network structure, the optimal combination of dynamical process, resolution value $n$ and (finite) horizon $m$~\cite{Fortunato2007} can be found {\em a-posteriori}.

At last we remark that the framework we developed is general and can possibly be applied to other kinds of networks (\eg, bipartite graphs) or to detect overlapping communities. Another interesting research direction would be to compare Markov processes of different nature within the quality function.

{\em Acknowledgments.} We thank an anonymous referee for useful suggestions.


%

\end{document}